\documentclass[letterpaper]{article} 
\usepackage{aaai2026}  
\usepackage{times}  
\usepackage{helvet}  
\usepackage{courier}  
\usepackage[hyphens]{url}  
\usepackage{graphicx} 
\usepackage{natbib}  
\usepackage{caption} 
\usepackage{amsmath}
\usepackage{booktabs}
\usepackage{multirow}
\usepackage[most]{tcolorbox}
\usepackage[ruled,vlined]{algorithm2e}%

\newtcblisting{promptbox}{
  colframe=black,
  colback=white,
  boxrule=0.8pt,
  arc=3pt,
  left=1mm,
  right=1mm,
  top=1mm,
  bottom=1mm,
  listing only,
  listing options={
    basicstyle=\ttfamily\small, 
    breaklines=true,
    numbers=left
  }
}

\usepackage[utf8]{inputenc} 
\usepackage[T1]{fontenc}    
\usepackage{url}            
\usepackage{booktabs}       
\usepackage{amsfonts}       
\usepackage{graphicx}       
\usepackage{nicefrac}       
\usepackage{microtype}      
\usepackage{listings}
\usepackage{amssymb}
\usepackage{dblfloatfix}
\usepackage{newfloat}
\usepackage{listings}
\DeclareCaptionStyle{ruled}{labelfont=normalfont,labelsep=colon,strut=off} 
\title{TRAIL: Joint Inference and Refinement of Knowledge Graphs with Large Language Models}
\author{
    Xinkui Zhao\textsuperscript{\rm 1},
    Haode Li\textsuperscript{\rm 1},
    Yifan Zhang\textsuperscript{\rm 1},
    Guanjie Cheng\textsuperscript{\rm 1},
    Yueshen Xu\textsuperscript{\rm 2}
}
\affiliations{
    \textsuperscript{\rm 1}School of Software Technology, Zhejiang University\\
    \textsuperscript{\rm 2}School of Computer Science and Technology, Xidian University\\
}
\usepackage{bibentry}

\begin{document}

\maketitle
\begin{abstract}
Recent advances in large language models (LLMs) have unlocked powerful reasoning and decision-making capabilities. However, their inherent dependence on static parametric memory fundamentally limits their adaptability, factual accuracy, and interpretability in knowledge-intensive scenarios. Knowledge graphs (KGs), as structured repositories of explicit relational knowledge, offer a promising approach for augmenting LLMs with external, interpretable memory. Nevertheless, most existing methods that combine LLMs with KGs treat reasoning and knowledge updating as separate processes, resulting in suboptimal utilization of new information and hindering real-time updates.
In this work, we propose \textbf{TRAIL}: a novel, unified framework for \textbf{T}hinking, \textbf{R}easoning, \textbf{A}nd \textbf{I}ncremental \textbf{L}earning that couples joint inference and dynamic KG refinement with large language models. TRAIL enables LLM agents to iteratively explore, update, and refine knowledge graphs during the reasoning process, employing a confidence-driven mechanism for the generation, validation, and pruning of new facts. This plug-and-play architecture facilitates seamless integration with various LLMs, supporting continual adaptation without the need for retraining. Extensive experiments on multiple benchmarks demonstrate that TRAIL outperforms existing KG-augmented and retrieval-augmented LLM baselines by 3\% to 13\%. More importantly, these results represent a significant step toward developing adaptive, memory-augmented language models capable of continual learning and reliable, transparent reasoning.
\end{abstract}

\section{Introduction}
The rapid advancement of large language models (LLMs) has transformed capabilities across a wide range of domains, including natural language processing~\cite{achiam2023gpt, touvron2023llama, liu2024deepseek} and code generation~\cite{guo2024deepseek, li2023starcoder, zhu2024deepseek, chowdhery2023palm, fried2022incoder}. Despite their remarkable performance, mainstream LLMs are fundamentally limited by their reliance on static parametric memory, which renders them susceptible to hallucination~\cite{ji2023survey,katz2024gpt,onoe2022entity}, knowledge staleness~\cite{chu2025empirical,wang2024memoryllm}, and a lack of interpretability—challenges that are particularly pronounced in knowledge-intensive domains~\cite{yin2022survey}. To address these limitations, knowledge graphs (KGs) have emerged as a powerful form of structured external memory, offering explicit, relational representations of real-world facts that can be systematically accessed and interpreted~\cite{edge2024local,liang2025kag,peng2024graph,procko2024graph}. Integrating LLMs with KGs has demonstrated promise in improving factual accuracy, enabling multi-hop reasoning, and enhancing transparency, thereby facilitating the development of more trustworthy and controllable AI systems.

\begin{figure*}
    \centering
    \includegraphics[width=1\linewidth]{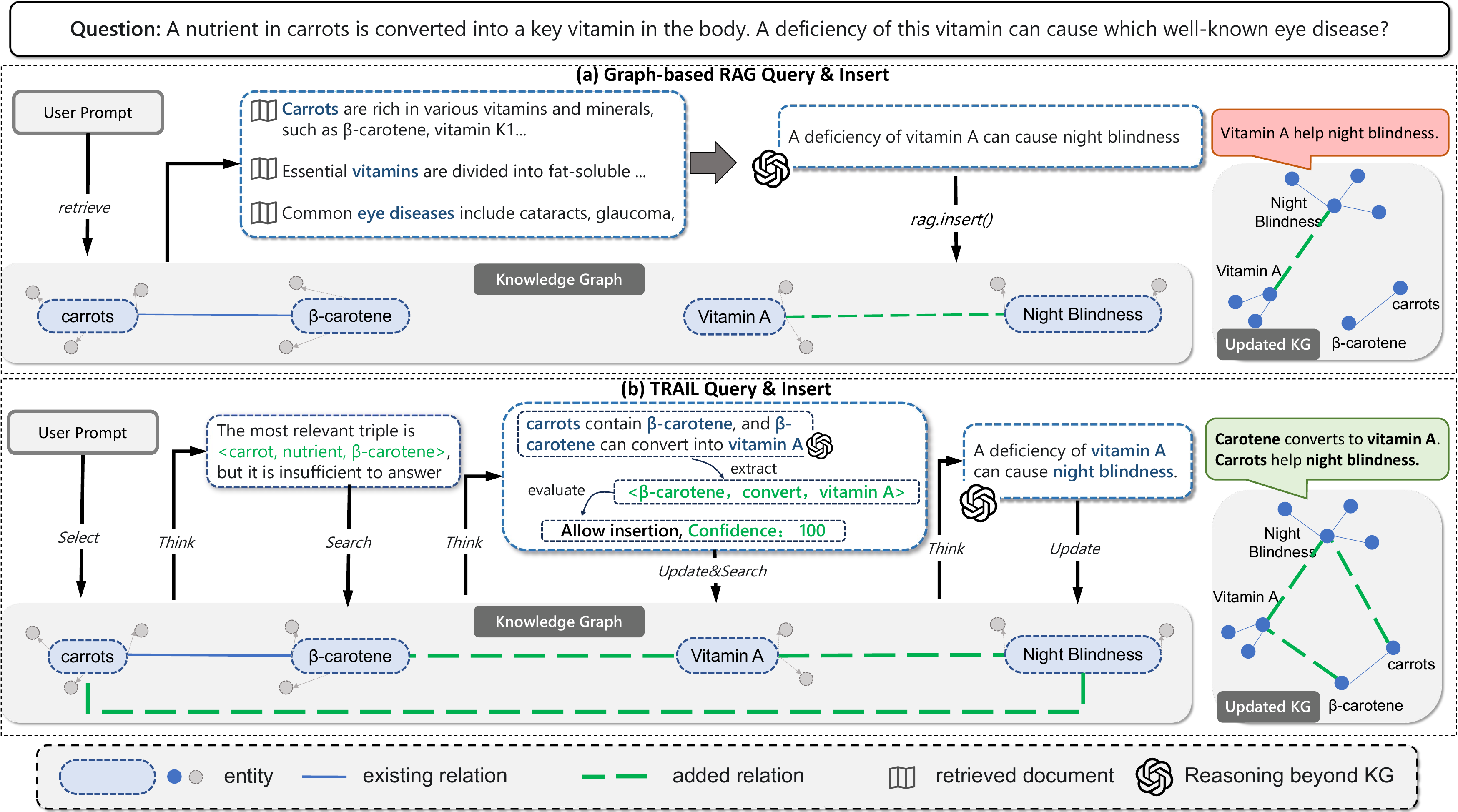}
    \caption{(a) Graph-based RAG Query \& Insertion:Related entities are retrieved and provided to the LLM to generate an answer. KG is updated after the answer is produced. (b) TRAIL Query \& Insertion: KG is updated dynamically during the reasoning process, allowing more information to be retained.}
    \label{fig:trail}
\end{figure*}

However, existing approaches to KG-augmented LLM reasoning typically treat the knowledge base as a fixed, read-only resource~\cite{luo2023reasoning,xu2024generate}, decoupling the processes of knowledge updating and logical inference. As illustrated in Figure~\ref{fig:trail}(a), the graph-based retrieval-augmented generation (RAG) approach separates retrieval and insertion: relevant entities are first retrieved from the knowledge graph and provided to the LLM for answer generation. If the LLM possesses sufficient knowledge, it can generate accurate answers even with an incomplete graph. Knowledge graph updates occur only after answer generation and usually involve adding a small number of new entities or relations related to the response. This separation prevents the model from dynamically refining or accumulating knowledge during reasoning, resulting in newly inferred information often being omitted from the knowledge graph. Consequently, the system struggles to adapt to new evidence or bridge knowledge gaps in real time. This deficiency is particularly pronounced in applications such as personal agent assistants, where the knowledge graph must serve as an up-to-date, continuously evolving repository of personal information.

To address these challenges, we propose \textbf{TRAIL}: a novel framework that unifies \textbf{T}hinking, \textbf{R}easoning, \textbf{A}nd \textbf{I}ncremental \textbf{L}earning for joint inference and knowledge graph refinement with large language models. Unlike conventional approaches that treat knowledge graphs as static, read-only resources, TRAIL enables the LLM agent to interactively traverse, expand, and refine the knowledge graph throughout the reasoning process as shown in Figure~\ref{fig:trail}(b). At each step, the agent dynamically decides whether to retrieve relevant subgraphs, hypothesize missing facts, or synthesize an answer, guided by both the evolving context and a confidence-based evaluation mechanism. This tight integration of reasoning and knowledge acquisition empowers TRAIL to incrementally accumulate new facts, resolve inconsistencies, and correct outdated information on-the-fly—without the need for retraining or manual intervention.

Crucially, TRAIL is designed as a \textbf{modular}, \textbf{plug-and-play} architecture that can be seamlessly integrated with a wide range of LLMs and downstream tasks. By coupling inference with continual knowledge graph refinement, TRAIL bridges the gap between static knowledge and adaptive reasoning, enabling the development of memory-augmented language models capable of lifelong learning. As demonstrated in our experiments, this framework not only delivers improved factual accuracy and interpretability compared to existing KG-augmented LLMs, but also provides a versatile foundation for building trustworthy and self-improving AI systems.

Our work makes the following key contributions:

\begin{itemize}
    \item We propose \textbf{TRAIL}, a unified framework that tightly integrates thinking, reasoning, and incremental learning, enabling large language models to jointly perform inference and dynamic knowledge graph refinement in an interactive manner.
    \item We introduce a confidence-driven mechanism for real-time validation, refinement, and pruning of newly generated knowledge, ensuring both robustness and trustworthiness of the evolving knowledge graph during reasoning.
    \item We design a modular, plug-and-play architecture that can be seamlessly combined with diverse LLMs and applied to a wide range of domains, supporting continual adaptation and knowledge transfer without additional retraining.
    \item Through comprehensive experiments on multiple medical QA benchmarks, we demonstrate that TRAIL consistently outperforms traditional KG-augmented and retrieval-augmented LLM baselines in both accuracy and interpretability, advancing the development of adaptive, memory-augmented language models.
\end{itemize}

\section{Background}

Despite the remarkable capabilities of large language models (LLMs), mainstream LLMs are fundamentally constrained by their static parametric memory. This limitation renders them susceptible to hallucinations~\cite{ji2023survey,katz2024gpt,onoe2022entity}, knowledge staleness~\cite{chu2025empirical,wang2024memoryllm}, and a lack of interpretability, particularly in knowledge-intensive domains~\cite{yin2022survey}. To address these shortcomings, external knowledge retrieval~\cite{kobayashi2000information,singhal2001modern} has emerged as a practical solution, providing reliable and up-to-date information for knowledge-intensive or domain-specific applications. Retrieval methods have also been effectively incorporated into advanced generative models in the era of AI-Generated Content (AIGC)\cite{li2024empowering,wu2024coral,sheynin2022knn}. Among these, Retrieval-Augmented Generation (RAG)\cite{lewis2020retrieval} stands out as one of the most prominent paradigms. RAG leverages information from external data sources as supplementary references or instructions for both input queries and generated outputs~\cite{min2020ambigqa,fan2024survey,hu2024rag}.

However, vanilla RAG approaches often suffer from two key limitations: the lack of global information and the neglect of relational structure, both of which hinder their performance on complex reasoning tasks. To address these challenges, Graph Retrieval-Augmented Generation (GraphRAG)~\cite{edge2024local,hu2024graggraphretrievalaugmentedgeneration} has been proposed. Unlike traditional RAG, GraphRAG retrieves graph elements containing relational knowledge relevant to the query from a pre-constructed knowledge graph database. 
While GraphRAG and similar methods enhance LLM reasoning by providing structured relational information from static knowledge graphs, they still exhibit limitations due to loose coupling between the LLM and KG, and their reliance on the completeness of the underlying KG. Most existing paradigms simply use the LLM to translate user queries into KG queries, treating the actual reasoning over the KG as an independent process. Consequently, when the KG is incomplete or lacks certain relations, these approaches may still fail to answer knowledge-intensive or multi-hop questions.

To address these issues, a general and increasingly prevalent paradigm for reasoning over knowledge graphs with LLMs is to instantiate the LLM as an autonomous agent that interacts with the KG in a stepwise manner~\cite{sun2023think,xu2024generate}. Within this framework, the agent continually aligns its actions with the original query to maintain goal relevance and dynamically assesses its current memory state to evaluate informational sufficiency and determine the most effective next step. At each iteration, the agent strategically decides whether to explore new entities and relations from the KG, hypothesize missing facts based on its parametric knowledge, or synthesize a final answer once sufficient evidence has been gathered. This agent-based approach enables flexible, interpretable, and robust multi-hop reasoning, and forms the foundation of many recent advancements in LLM-augmented KG reasoning.

\section{Motivation}
While the integration of LLMs with KGs has notably enhanced factual accuracy and interpretability, a critical limitation still remains: KGs are typically treated as static, read-only repositories, and even verifiable knowledge newly inferred during reasoning is seldom fed back into the graph. This disconnect gives rise to several key challenges:

\begin{itemize}
\item \textbf{Ephemeral Knowledge Acquisition}: Valuable intermediate inferences generated during reasoning are often discarded, resulting in repeated work and missed opportunities for knowledge accumulation.
\item \textbf{Limited Adaptivity and Coverage}: Separating reasoning from KG refinement limits the system’s capacity to incorporate new evidence, fill knowledge gaps, and mitigate the effects of outdated information in the KG.
\item \textbf{Barriers to Trustworthy and Lifelong Learning}: Without mechanisms for real-time validation and integration of new knowledge, systems cannot evolve or improve over time, limiting their trustworthiness, transparency, and potential for continual self-improvement.
\end{itemize}

To address these challenges, a unified framework is needed that supports both real-time reasoning and knowledge graph refinement. Such a framework should allow LLM agents to validate, integrate, and write back new knowledge during inference, enabling continual accumulation and refinement of facts throughout the reasoning process.

\section{Methodology}

\begin{figure}
    \centering
    \includegraphics[width=1\linewidth]{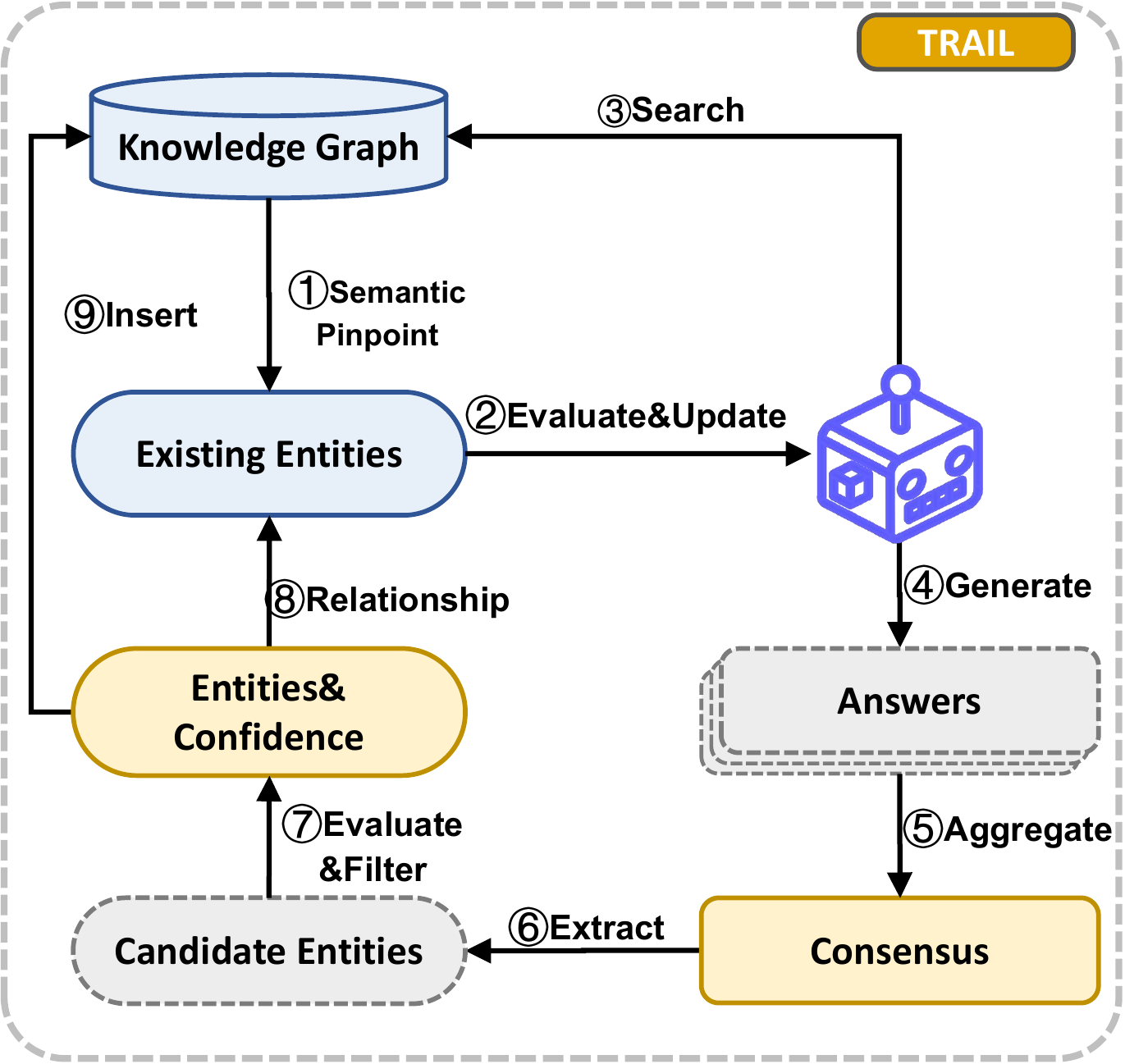}
    \caption{Schematic of the \textbf{TRAIL} framework. The agent iteratively retrieves and updates entities from the knowledge graph, assigns confidence scores, and validates new information through aggregation and consensus before integration. This process enables real-time answer synthesis and continual knowledge graph refinement.}
    \label{fig:method}
\end{figure}

In this section, we introduce our unified framework, \textbf{TRAIL} that tightly integrates \textbf{T}hinking, \textbf{R}easoning, \textbf{A}nd \textbf{I}ncremental \textbf{L}earning. TRAIL seamlessly couples large language models with knowledge graphs to enable joint inference and dynamic KG refinement. In contrast to existing approaches that treat the agent solely as a traverser of the KG, TRAIL empowers the agent to incrementally generate, validate, and integrate new facts into the KG throughout the reasoning process, as illustrated in Figure~\ref{fig:method}. This capability facilitates more effective and adaptive multi-hop reasoning, particularly when dealing with incomplete or evolving knowledge graphs.
\subsection{Problem Formulation}

Let $\mathcal{G} = \{(h, r, t)\}$ denote a knowledge graph, where $h$ and $t$ are entities and $r$ is a relation. Given a user query $q$, the goal is to predict the answer entity $a$, which may require multi-hop reasoning over $\mathcal{G}$, to leverage the new, verified knowledge generated during this process to refine $\mathcal{G}$ for future queries.

\subsection{Seed Point Selection}
Unlike existing KGQA benchmarks where the seed entity or region for reasoning is typically pre-specified and unambiguous~\cite{sun2023think,xu2024generate}, open-domain question answering scenarios present a significant challenge: the starting point for reasoning is often highly uncertain and lacks explicit guidance. However, methods that rely solely on semantic embedding similarity for seed entity selection, such as GraphRAG~\cite{edge2024local}, are susceptible to several limitations. These approaches often exhibit a bias toward high-degree hub entities or densely connected supernodes, as such nodes tend to appear semantically similar to a broad spectrum of queries in the embedding space. This bias introduces substantial irrelevant or noisy context into the reasoning process and can cause uncontrolled expansion of the retrieved subgraph.
To address these challenges, we propose a \textit{multi-stage seed point selection framework} that leverages the dual-role capability of LLMs:

\begin{enumerate}
    \item \textbf{Topic Identification}: Given query $q$, the LLM extracts a set of core semantic topics $\mathcal{T}$ based on contextual understanding rather than keyword matching.
    \item \textbf{Entity Anchoring}: Each topic $t_j$ is embedded and its Top-$K$ most similar entities $\mathcal{V}_{\mathrm{cand}}^{(j)}$ are retrieved from the entity embedding index, ensuring both coverage and diversity:
    \begin{equation}
        \mathcal{V}_{\mathrm{cand}}^{(j)} = \operatorname{TopK}_{v_i \in \mathcal{V}} S_{\mathrm{sim}}(t_j, v_i)
    \end{equation}
    \item \textbf{Heuristic Selection}: All candidates $\mathcal{V}_{\mathrm{cand}} = \bigcup_j \mathcal{V}_{\mathrm{cand}}^{(j)}$ and topic context are input to the LLM, which selects the final seeds $\mathcal{V}_{\mathrm{seed}}$ considering semantic relevance, balanced connectivity, and robustness for downstream inference:
    \begin{equation}
        \mathcal{V}_{\mathrm{seed}} = \operatorname{LLMSelect}(\mathcal{V}_{\mathrm{cand}}, \mathcal{T}, q)
    \end{equation}
\end{enumerate}

\subsection{Knowledge Graph Refinement}

To enable real-time adaptability and maintain the reliability of the evolving knowledge graph during reasoning, TRAIL incorporates a \emph{KG refinement} module. This component is essential for supporting incremental learning in open-domain and user-interactive scenarios, where the incompleteness and continuous evolution of knowledge are inherent challenges.

The KG refinement process is structured around two key operations: (1) \textbf{Confidence Evaluation}, (2) \textbf{Evolving Generated Entities and Relations}. 

\subsubsection{Confidence Evaluation}
Given the inherent limitations of LLMs like hallucination and inconsistency, ensuring the reliability of our KG is paramount, especially as it becomes the foundation for subsequent reasoning steps. Unconditionally accepting all outputs from the reasoning process may propagate errors and compromise the integrity of the knowledge base. To address this, we introduce a \textit{confidence evaluation mechanism} for new nodes.

Formally, we assign to each triple $(h, r, t)$ in the knowledge graph a confidence score $c(h, r, t) \in [0, 100]$ as follows:
\[
c(h, r, t) = 
\begin{cases}
100, & \text{if } (h, r, t) \in \mathcal{G}_\mathrm{truth} \\
\mathrm{JudgeLLM}(h, r, t), & \text{if } (h, r, t) \text{ is generated}
\end{cases}
\]
where $\mathrm{JudgeLLM}(\cdot)$ denotes the output of the evaluation LLM, normalized to $[0, 100]$. Specially, we employ a dedicated evaluation LLM—distinct in architecture and training data from the reasoning model—is prompted to assess the plausibility and reliability of each candidate fact. This separation ensures fair and unbiased scoring. The evaluation process uses stringent prompts, designed to elicit critical and rigorous judgment from the evaluation model.

\begin{table*}[h]
\centering
\footnotesize
\setlength{\tabcolsep}{4pt} 
\renewcommand{\arraystretch}{1.2}
\caption{\textbf{Effectiveness of \emph{TRAIL}.} The best best is highlighted in red. Our method is listed at the bottom and is presented in \textbf{bold} for emphasis.}
\begin{tabular}{lccccc}
\toprule
\textbf{Models} & \textbf{MedQA} & \textbf{MedMCQA} & \textbf{PubMedQA} & \textbf{MMLU-Pro\_Health} & \textbf{MMLU-Pro\_Biology} \\
\midrule
\multicolumn{6}{c}{\textit{Comparative Models}} \\
DeepSeek-V3      & 79.6 & 72.0 & 82.2 & 72.6 & 84.8 \\
Llama-3.1-70B-Instruct & 78.4 & 72.5 & 78.5 & 68.2 & 80.8 \\
Qwen2.5-72B-Instruct & 72.7 & 66.2 & 71.7 & 65.3 & 78.8 \\
HuatuoGPT-o1-70B    & 83.3 & 73.6 & 80.6 & 71.0 & 82.8 \\
HuatuoGPT-o1-72B      & 81.4 & 73.6 & 84.9 & 72.0 & 82.8 \\
\multicolumn{6}{c}{\textit{Comparative Methods}} \\
LightRAG (global)           & 82.1 & 70.6 & 82.0 & 69.2 & 85.7 \\
LightRAG (local)           &82.0 & 67.1 & 82.5  & 67.7 & 82.7  \\
ToG     & 79.9 & 69.4 & 80.6 & 70.2 & 83.8\\
\midrule
\multicolumn{6}{c}{\textit{Ours}} \\
\textbf{TRAIL}          & \textbf{79.9} & \textbf{72.1} & \textbf{82.6} & \textbf{76.5} & \textbf{88.7} \\
\bottomrule
\end{tabular}%

\label{tab:main_results}
\end{table*}

\subsubsection{Evolving Generated Entities and Relations}
The evolution of generated entities and relations in our framework involves two key components: \textit{Insertion} and \textit{Refinement}. The former adds generated entities and relations to KG, while the latter adjusts or prunes the previously inserted entities and relations to improve consistency and accuracy. The entire process is illustrated in Algorithm~\ref{alg:entity}.

\begin{algorithm}[ht]
\caption{Evolving Generated Entities and Relations }
\KwIn{Current KG $\mathcal{G}$, Query $Q$, Confidence Threshold $\tau$}
\KwOut{Updated KG $\mathcal{G}'$}

\While{Reasoning is ongoing} {
    \If{Dead-end reached}{
         Generate candidate outputs;\\
        Aggregate outputs to consensus;\\
        Parse consensus to entities/relations;\\
        \ForEach{entities/relations}{
            Evaluate confidence\;
            \If{confidence $> \tau$}{
                Insert into $\mathcal{G}$\;
            }
        }
    }
    
    \ElseIf{Expanding generated node}{
        Re-evaluate confidence;\\
        Compute new confidence;\\
        \If{new confidence $< \tau$}{
            Prune\;
        }
        Refinement for reasoning\;
    }
    Update session cache\;
}
\Return{Updated KG $\mathcal{G}'$}
\label{alg:entity}
\end{algorithm}

\textbf{Insertion.} Our framework automatically adds only verified and evaluated entities and relations during the reasoning process to address the incompleteness of static knowledge graphs, allowing for the hypothesis and incorporation of novel facts when explicit information is unavailable. This mechanism is triggered during the \textit{Generate} phase when the reasoning process reaches a dead-end —— that is, when search fails to discover a viable path in the KG.

To enhance factual accuracy and reduce randomness in the generated candidates, we set the LLM’s decoding temperature to a lower value during generation, thereby making its outputs more deterministic and reliable. To ensure quality and prevent noise, we adopt aggregation strategies inspired by recent advances such as MoA~\cite{wang2024mixture} and LightRouter~\cite{zhang2025lightrouter}. We sample multiple candidate outputs at each generation step and employ a secondary model to aggregate them, reaching a consensus among the candidates. This aggregation mechanism enhances the robustness and quality of the generated facts.

The aggregated consensus output is then parsed to extract entities and relations, which are organized into a candidate list for potential integration into the knowledge graph. Each candidate entity is evaluated using the previously described confidence evaluation mechanism. Only those candidates that surpass a predefined confidence threshold are inserted into the graph, with their confidence scores set according to the evaluation results.

\textbf{Refinement.} Our framework continuously refines entities and relations to prevent knowledge decay and ensure adaptability in a changing information landscape. Re-evaluation of hypothesized facts occurs either when new evidence emerges or when the reasoning model is upgraded to a more capable variant. This mechanism is initiated during the \textit{Search} phase: as reasoning expands to neighboring nodes added in prior \textit{Generate} phase, confidence re-evaluation is triggered to verify their validity and relevance, enabling dynamic refinement.

When the evaluation determines that refinements are necessary—such as updating entity descriptions or modifying inter-entity relations—the confidence score is recalculated by combining the previous score with the new evaluation to reflect the latest trust level. If this updated confidence falls below a set threshold, the entity and its connected edges are pruned from the graph to maintain data integrity.

To improve efficiency and prevent redundant or infinite re-evaluations, a session-specific cache ensures that each node is rescored at most once per reasoning session. This dynamic refinement and re-evaluation keeps the knowledge graph adaptive and reliable, integrating new information while preserving the integrity of generated entities. The updated node and its revised confidence score are then used in subsequent reasoning steps, allowing the system to leverage the most current and trustworthy knowledge.




\section{Experiment Setup}

\paragraph{Truth KG Construction}
Unlike conventional KGQA settings that assume a complete knowledge graph, we focus on a more realistic scenario where the space of real-world questions far exceeds the knowledge explicitly encoded in the KG. To simulate this setting, we construct an incomplete knowledge graph from a medical dataset~\cite{chen2024huatuogpto1medicalcomplexreasoning} containing 40.6k factual records. The first 30k records are used to build the KG using LightRAG~\cite{guo2024lightrag}.

\paragraph{Dataset}
For a more comprehensive and detailed evaluation of our framework, we select five widely-used benchmark datasets in the medical domain: MedQA~\cite{jin2020disease}, MedMCQA~\cite{pmlr-v174-pal22a}, PubMedQA~\cite{jin2019pubmedqa}, MMLU-Pro\_Health~\cite{wang2024mmlu}, and MMLU-Pro\_Biology.

\paragraph{TRAIL Setup}
Deepseek-V3~\cite{liu2024deepseek} is employed as the reasoning model and Deepseek-R1~\cite{guo2025deepseek} is employed as the evaluation model.

\paragraph{Methods Selected for Comparison}
To comprehensively evaluate the effectiveness of our proposed TRAIL framework, we compare it with several strong and representative baselines across multiple benchmarks. Specifically, we consider the following categories:

\textbf{Comparative Models}: We select leading open-source LLMs, including DeepSeek-V3~\cite{liu2024deepseek}, Llama-3.1-70B-Instruct~\cite{touvron2023llama}, Qwen2.5-72B-Instruct~\cite{team2024qwen2}, HuatuoGPT-o1-70B, and HuatuoGPT-o1-72B~\cite{chen2024huatuogpto1medicalcomplexreasoning}. These models are evaluated in a zero-shot setting without explicit access to external structured knowledge, serving as pure LLM baselines.

\textbf{Retrieval-Augmented Generation}: We implement LightRAG~\cite{guo2024lightrag} in both global and local retrieval modes to benchmark the effect of integrating external knowledge graphs with LLMs. LightRAG (Global) retrieves relevant information from the entire graph, while LightRAG (Local) focuses on contextually local subgraphs. Deepseek-V3~\cite{liu2024deepseek} is employed as the reasoning model.

\textbf{Graph-based Reasoning}: ToG~\cite{sun2023think} is included as a representative graph-based reasoning baseline, enabling multi-hop inference over knowledge graphs. Deepseek-V3~\cite{liu2024deepseek} is employed as the reasoning model.

\section{Experiment Result}
Table~\ref{tab:main_results} presents the performance of TRAIL and several baseline models across five medical and biological question answering benchmarks. Our proposed method, \textbf{TRAIL}, consistently achieves strong results, outperforming all baseline methods on four out of five datasets.

Notably, TRAIL attains the highest accuracy on both MMLU-Pro\_Health (76.5\%) and MMLU-Pro\_Biology (88.7\%). This achievement is particularly significant, as the MMLU-Pro dataset is specifically designed to evaluate deep, multi-hop reasoning abilities rather than simple fact retrieval. The questions in this benchmark require synthesizing multiple, disparate pieces of knowledge to form coherent logical chains. Thus, TRAIL's superior performance on these tasks highlights its effectiveness in addressing complex reasoning challenges. Compared to LightRAG and ToG, TRAIL demonstrates clear improvements, especially in domains that demand continual knowledge accumulation and adaptive inference. These results underscore the strengths of our unified framework, which enables LLMs not only to leverage but also to incrementally refine the knowledge graph throughout the reasoning process.

\section{Discussion}

\subsection{Does Reasoning Model Ability Matter?}

\begin{figure}[t]
    \centering
    \includegraphics[width=1\linewidth]{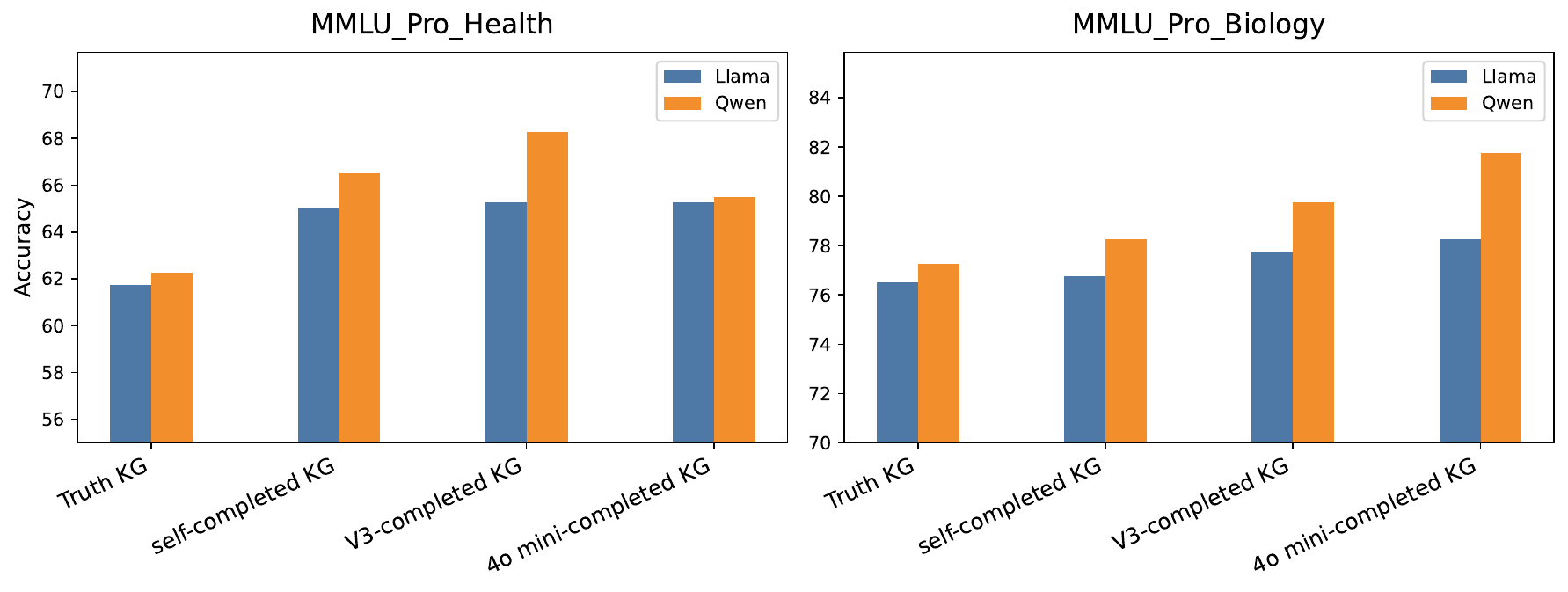}
    \caption{Accuracy of KG refinement strategies on MMLU-Pro\_Health and MMLU-Pro\_Biology. "Truth KG" is the original incomplete graph; other KGs are completed by self, V3, or 4o mini models. Results are reported for both Llama and Qwen.}
    \label{fig:reason}
\end{figure}

As shown in Figure~\ref{fig:reason}, we assess the effects of various KG refinement strategies on MMLU-Pro\_Health and MMLU-Pro\_Biology. The results demonstrate that the baseline Truth KG yields the lowest accuracy for both Llama-3.1-70B-Instruct(hereafter Llama in figures and tables) and Qwen2.5-72B-Instruct(hereafter Qwen in figures and tables) as reasoning model, indicating that KG incompleteness is a major bottleneck for downstream performance.

Notably, even basic self-refinement of the knowledge graph consistently improves accuracy, underscoring the importance of addressing missing information. Moreover, employing more advanced refinement models—such as DeepSeek-V3 (hereafter referred to as "V3" in figures and tables) and GPT-4o mini (hereafter referred to as "4o mini" in figures and tables)—yields substantial and consistent performance gains, most notably on MMLU-Pro\_Biology with the Qwen agent. These findings highlight that the capability of the underlying refinement model is critical to downstream performance. Across all settings, Qwen slightly outperforms Llama, though both exhibit the same trend: improvements in KG quality directly translate to enhanced task performance. By utilizing moderate-capacity retrieval agents, we ensure that the observed gains are primarily attributable to improvements in KG quality rather than the capabilities of the reasoning model itself.

\subsection{Can KG be Continuously Improved?}
\begin{figure}[t]
    \centering
    \includegraphics[width=1\linewidth]{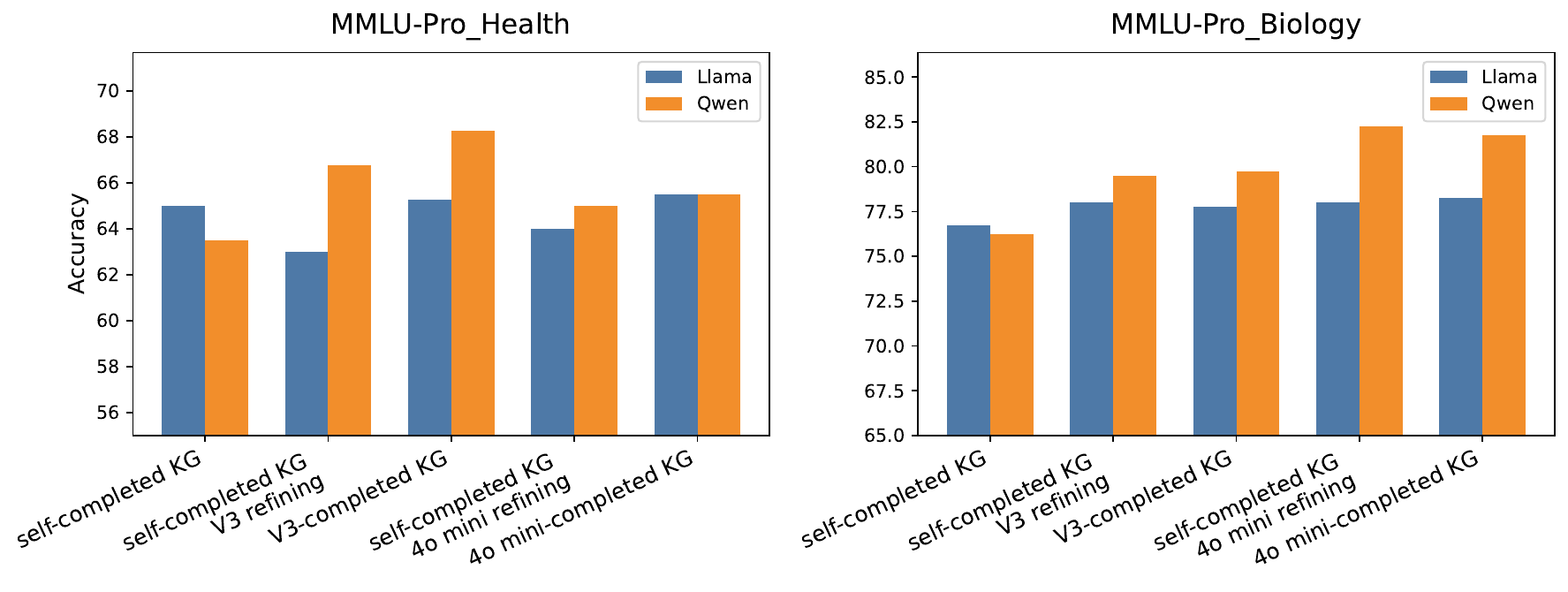}
    \caption{Performance of different KG refinement strategies on MMLU-Pro\_Health and MMLU-Pro\_Biology. “A-completed KG” uses KGs completed by model A; “A-completed KG B refining” further refines these with model B. Results are shown for both Llama and Qwen as reasoning models.}
    \label{fig:edit}
\end{figure}

The results in Figure~\ref{fig:edit} clearly demonstrate that knowledge graph refinement significantly enhances performance both in MMLU-Pro\_Health and MMLU-Pro\_Biology. The data reveals that an initially underperforming KG can achieve a substantial performance leap after being refined by a more advanced model. This phenomenon strongly suggests that the refinement process is not merely a simple "fix" but rather an effective "evolution" mechanism. It indicates that a flawed initial KG does not need to be discarded; instead, it can serve as a foundation for iterative optimization by integrating it with more powerful models.

\subsection{Generalizability of TRAIL}
Building on the previous experiments, which demonstrate the generalizability of TRAIL across different tasks and models, we further evaluate the utility of the constructed knowledge graphs when used with the LightRAG retriever. As shown in Table~\ref{tab:rag-table}, incorporating completed KGs leads to substantial performance improvements for LightRAG on both MMLU-Pro\_Health and MMLU-Pro\_Biology.

Specifically, LightRAG achieves the lowest accuracy when using the original Truth-KG, while refinement with stronger models (Deepseek-V3 and GPT-4o mini) consistently boosts accuracy. These results further confirm that high-quality, completed knowledge graphs are broadly beneficial and can be effectively leveraged by different RAG frameworks.

\begin{table}[ht]
\centering
\caption{MMLU Pro Health and Biology Benchmark Results with LightRAG using Completed KG}
\label{tab:rag-table}
\resizebox{\columnwidth}{!}{
\begin{tabular}{lcccc}
\toprule
\multirow{2}{*}{Benchmark} & \multirow{2}{*}{Qwen} & \multicolumn{3}{c}{LightRAG} \\
\cmidrule(l){3-5}
 & & Truth-KG & V3-KG & 4o-mini-KG \\
\midrule
MMLU-Pro\_Health & 65.3 & 66.7 & 73.6 & 73.4 \\
MMLU-Pro\_Biology & 78.8 & 81.5 & 82.0 & 84.2 \\
\bottomrule
\end{tabular}
} 
\end{table}

\subsection{A Deeper Discussion: Applications}

\begin{figure}
    \centering
    \includegraphics[width=1\linewidth]{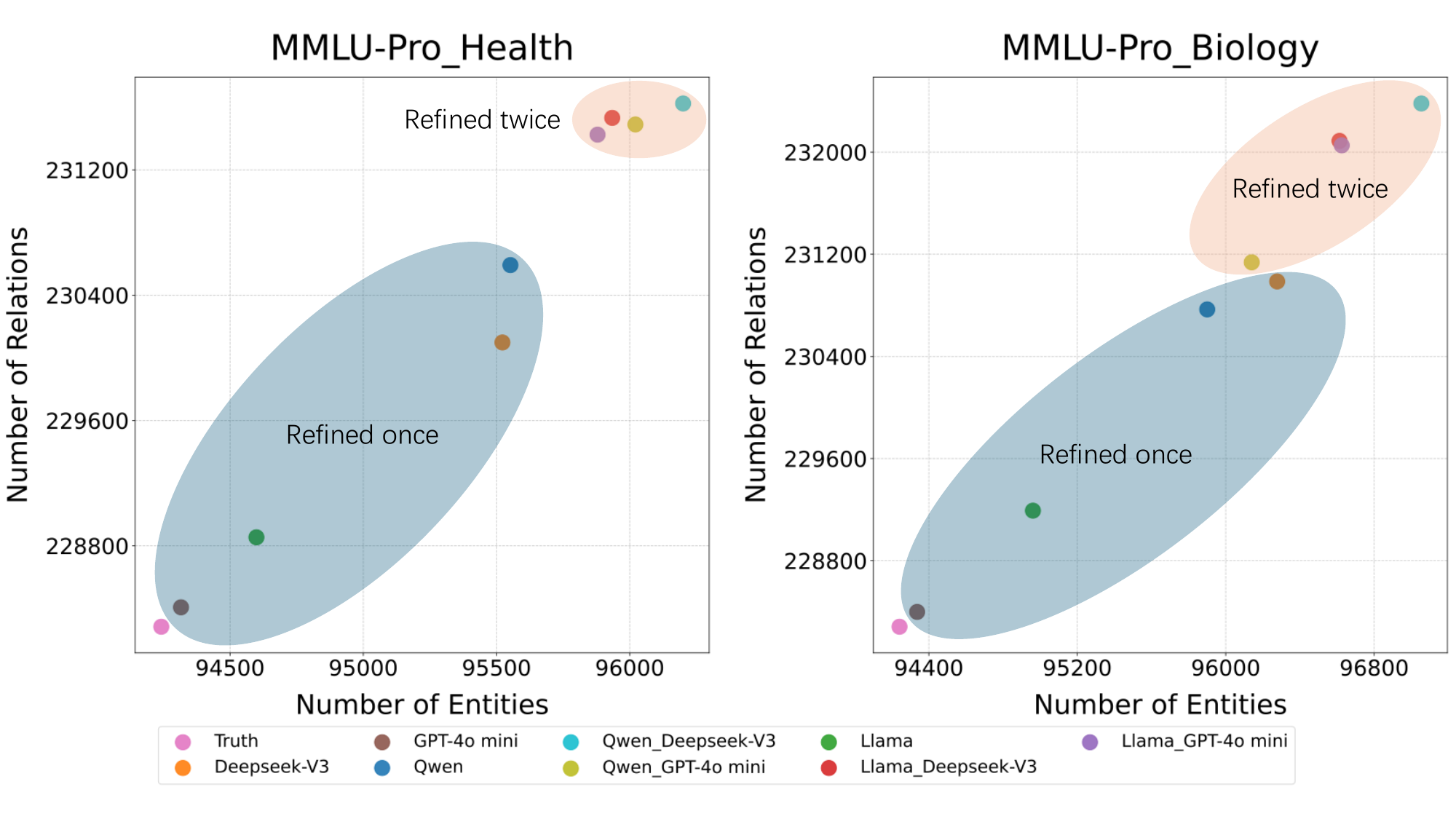}
    \caption{KG size after different TRAIL configurations.}
    \label{fig:kgsize}
\end{figure}

Our framework flexibly integrates with LLMs of different scales, enabling the construction of enriched knowledge graphs (as shown in Figure~\ref{fig:kgsize}) as structured external memories. By using strong LLMs for KG refinement and reasoning, we provide high-quality KGs that lightweight models can efficiently use for reasoning and retrieval, supporting deployment in resource-limited settings. As LLMs improve, our framework supports iterative KG updates without full reconstruction, ensuring the KG remains accurate and current. The modular design allows different models to be used for KG construction, validation, and inference, and makes it easy to transfer or share KGs across tasks and domains, improving both scalability and flexibility.

These KGs can be easily transferred between models and retrieval methods. Future work may explore dynamic switching between RAG and TRAIL to balance efficiency and cost. When used as agent memory, the KG also supports integration of personalized or user-specific knowledge, enabling more context-aware and adaptive behavior.

\section{Related Work}
\subsection{Reasoning with LLM}
LLMs have achieved strong results in natural language tasks, yet complex reasoning remains a challenge. Early models like GPT-3~\cite{brown2020language} excelled at language understanding but struggled with reasoning, prompting the development of advanced prompting and training methods. Chain-of-thought and multi-path prompting~\cite{wei2022chain,zhang2022automatic,kojima2022large,yao2023tree,besta2024graph} significantly improved reasoning by decomposing tasks and enabling exploration of diverse solution paths. Beyond prompting, techniques such as reinforcement learning (e.g., RLHF, DPO)\cite{ouyang2022training,rafailov2023direct}, step-level supervision\cite{chen2024step}, and search-based strategies (e.g., MCTS)\cite{silver2016mastering,wang2022self} further align LLM behavior with human reasoning. Recent models\cite{guo2025deepseek,zhong2024evaluation} advance the field by integrating both outcome-based and process-based supervision with sophisticated search at inference time.

\subsection{KG-Enhanced LLM Reasoning}
Integrating KGs with LLMs has emerged as a promising approach to enhance factual accuracy, reasoning depth, and reliability in question answering and complex reasoning tasks. Recently, retrieval-augmented methods such as GraphRAG~\cite{edge2024local,liang2025kag,procko2024graph,guo2024lightrag} have established a new paradigm for customizing LLMs with well-structured external knowledge and improved contextual reasoning, and have demonstrated success across various domains~\cite{chen2025academicrag,hang2025trumorgpt,du2024codegrag}. A notable line of research in this area treats the LLM as an intelligent agent that can interactively explore and reason over KGs. For instance, ToG~\cite{sun2023think} systematically formulates LLM-based multi-hop reasoning as an agentic path exploration process on the KG. Similarly, RoG~\cite{luo2023reasoning} first generates relation paths as faithful reasoning plans, and then uses them to retrieve valid reasoning chains from KGs for downstream inference. To further address the challenge of incomplete KGs, GoG~\cite{xu2024generate} empowers LLMs to dynamically generate missing factual triples during the reasoning process, thereby enabling robust hybrid reasoning based on both retrieved and generated knowledge.

\section{Conclusion}
This paper presents TRAIL, a unified framework for joint inference and dynamic refinement of knowledge graphs with large language models. By tightly coupling reasoning and knowledge graph evolution, TRAIL enables LLM agents to incrementally accumulate, validate, and update factual knowledge throughout multi-hop reasoning. Experimental results on multiple medical benchmarks demonstrate that TRAIL achieves superior accuracy and interpretability compared to existing KG-augmented and retrieval-augmented baselines. The proposed confidence-driven mechanism ensures the trustworthiness of the evolving knowledge base, while the modular architecture facilitates seamless integration with diverse LLMs and downstream applications. TRAIL provides a scalable foundation for developing adaptive, memory-augmented language models capable of continual learning, robust reasoning, and transparent decision-making.
\section{Limitations}
Despite the demonstrated effectiveness of TRAIL, several limitations remain. First, the overall performance of the framework is sensitive to the instruction-following and reasoning capabilities of the underlying language model; when deployed with less-aligned models, the accuracy of both KG refinement and multi-hop inference degrade noticeably. Second, TRAIL exhibits limited effectiveness in zero-shot and few-shot scenarios, where the model lacks sufficient context or domain adaptation. Additionally, the confidence evaluation and refinement process introduces computational overhead, which may affect scalability in real-time or resource-constrained environments.

\newpage


\bibliography{aaai2026}

\clearpage

\end{document}